\documentclass[12pt,iop]{emulateapj}
\usepackage{url}
\usepackage{rotating}
\usepackage{natbib}
\usepackage{xcolor}
\usepackage{longtable}

\newcommand{\clean}{{\sc clean}}

\shorttitle{MWA32 PAPER32 Comparison}
\shortauthors{Jacobs et al.}

\begin{document}

\title{The precision and accuracy of early Epoch of Reionization
  foreground models: comparing MWA and PAPER 32-antenna source
  catalogs}


\author{Daniel C. Jacobs\altaffilmark{1,2}}
\author{Judd Bowman\altaffilmark{1}}
\author{James E. Aguirre\altaffilmark{3}}

\altaffiltext{1}{School of Space and Earth Exploration, Arizona State University, Tempe, AZ} 
\altaffiltext{2}{Corresponding Author: \email{daniel.c.jacobs@asu.edu}}
\altaffiltext{3}{Department of Physics and Astronomy, University of Pennsylvania, 209 S. 33rd Street, Philadelphia, PA, USA}

\begin{abstract}

As observations of the Epoch of Reionization (EoR) in redshifted
21cm emission begin, we asses the accuracy of the early catalog 
results from the Precision Array for Probing the Epoch of Reionization
(PAPER) and the Murchison Widefield Array. The MWA EoR approach derives
much of its sensitivity from subtracting foregrounds to $<1$\%  precision  
while the PAPER approach relies on the stability and symmetry of the primary beam. 
Both require an accurate flux calibration to set the amplitude of the measured
power spectrum. The two instruments are very similar in resolution,
sensitivity, sky coverage and spectral range and have produced catalogs from nearly contemporaneous data. 
We use a Bayesian MCMC fitting method to estimate
that the two instruments are on the same flux scale to within 20\% and find
that the images are mostly in good agreement.  We then investigate
the source of the errors by comparing two overlapping MWA facets where we
find that the differences are primarily related to an inaccurate model of the primary beam
but also correlated errors in bright sources due to \clean. We conclude with 
suggestions for mitigating and better characterizing these effects.

\end{abstract}


\section{Introduction}

Recent interest in very high redshift ($6<z<12$) 21 cm HI emission
from the Epoch of Reionization
\citep[EoR, see reviews in][]{ Zaldarriaga:2004p2115,McQuinn:2006p2246,Furlanetto:2006p2267,Morales:2010p8093}
has inspired a renaissance of meter wavelength ($\nu<200$ MHz) radio
astronomy
Several telescopes, including the Giant
Metre-Wave Telescope
\citep[GMRT;][]{Swarup:1991p8685}\footnote{\url{http://gmrt.ncra.tifr.res.in/}},
the Low Frequency Array
\citep[LOFAR;][]{Rottgering:2006p9018}\footnote{\url{http://www.lofar.org}},
the Murchison Wide-field Array
\citep[MWA;][]{Tingay:2012p9022,Bowman:2012p9428}\footnote{\url{http://mwatelescope.edu}},
and the Precision Array for Probing the Epoch of Reionization
\citep[PAPER;][]{Parsons:2010p6757}\footnote{\url{http://eor.berkeley.edu}}
are beginning to characterize foregrounds and perform their first deep
integrations and set upper limits \citep{Paciga:2011p9470,Paciga:2013p9627}.  Both
PAPER and the MWA operate in the southern hemisphere, as will the
future Square Kilometer Arrays (SKA).

 The EoR signal will be a small spatial and spectral variation on top
 of bright foreground sources
 \citep{DiMatteo:2004p2230,Oh:2003p5097,Jelic:2008p2130,Bowman:2006p1887}.
 The separation of the EoR from these foregrounds is expected to be
 the dominant source of uncertainty and has been the focus of much
 study.  Though the spatial RMS of the unresolved background was
 initially calculated to be larger than the EoR signal
 \citep{DiMatteo:2002p2226}, later simulations found that the spectral
 smoothness of the unresolved background enabled accurate subtraction
 to acceptable levels in the $k$ modes of interest
 \citep{Morales:2004p2494,Morales:2006p1903,Wang:2006p2238,Jelic:2008p2130,Bowman:2009p7816,Datta:2010p8781,Chapman:2012p8505,Cho:2012p9385} or just by avoiding the contaminating modes entirely \citep{Parsons:2012p8896}.
 Simulations tackling parameter estimation, polarization and
 foreground subtraction all assume that all unresolved sources will be
 removed such that the errors are indistinguishable from the
 unresolved point source background both spatially and spectrally
 \citep{Liu:2011p8763,Bowman:2009p7816,Liu:2009p4762,Bowman:2009p7816,Harker:2010p6556,Gleser:2008p9387,Petrovic:2011p9386}.
 The level of source residual varies between these simulations. While
 \citet{Bowman:2009p7816} assume that subtraction will achieve a 10
 mJy residual, \citet{Liu:2009p4762} test a range of scenarios up to
 100 mJy residual flux.\footnote{The ultimate flux limit to which
   sources can be identified and removed is set by the resolution of
   the instrument. Bright extra-galactic point sources increase in
   number with decreasing brightness
   \citep{Condon:1998p7986,Lane:2008p7683,Baldwin:1985p5306,Hales:1988p9400,McGilchrist:1990p9402}. At
   some source flux level, the number of sources per synthesized beam
   becomes greater than one. For PAPER this limit is $\sim$100 mJy,
   while the MWA reaches to tens of mJy.} Since even the quietest
 fields of view contain several 40 Jy sources, these residual levels
 translate to removal precision requirements of $0.025$\% and $0.25$\%
 respectively! In contrast, most radio point source catalogs have flux
 accuracies in the 5 to 20\% range.  Studies of errors in bright
 source removal are limited. In one simulation that included bright
 source subtraction, \citet{Datta:2010p8781} found that point-source
 foregrounds extended further into the spectral dimension than were
 previously predicted into the so-called ``wedge".  This turns out to 
 be equivalent to the statement that longer baselines are contaminated at higher delays 
 which defines the Parsons et al ``wall" that defines the $k$ modes accessible to PAPER. 
 In both cases the implication is that the flux accuracy
 requirement extends to the spectral dimension.


The requirement of point source subtraction imposes accuracy
requirements in source modeling which have rarely been achieved in
practice.  According to estimates of catalog flux accuracy by
\citet[][reproduced in Table \ref{tab:surveys}]{Vollmer:2005p6425},
most catalog fluxes at high frequencies agree to within $\sim
5\%$. The study included only one catalog in the EoR band (4C), which
had by far the largest flux error (15\%).  Accordingly, attention has
begun to focus on approaches which largely avoid the need to model and
subtract sources to high accuracy \citep[for
  example,][]{Parsons:2012p8896}.  Even in the absence of a need for a
highly accurate sky model for EoR experiments, uncertainty in calibrator
flux translates directly into the overall amplitude of the power spectrum measurement
which limits the constraining power of the observation.
Reliable calibrators are also necessary for modeling the
instrument primary beam (which enters into the $k$-space window
function and noise estimates) and for generating reliable and
repeatable instrument calibrations.  For
example, attempts to model the primary beam are currently limited by
the accuracy to which source fluxes are known over a wide enough area
of sky to fully sample the beam \citep{Pober:2012p8800}.

%
\begin{deluxetable}{lrrrrrr}	
\tablecaption{SPECFIND radio continuum source catalogue entries and
  estimated uncertainty, adapted from \cite{Vollmer:2005p6425}}
\tablehead{
Survey  & $\nu_0$ & $\theta$ & $S_{min}$ & Source & Ref & Error \\
Name    & (MHz)     & (\arcmin)   & (mJy)    & count   &     & (\%)}
\startdata
\label{tab:surveys}
PMN  &  4850 &  3.5      &  20 & 50814  & 1 & 5\\
PKS  &  2700 &  8.0      &  50 & 8264  & 2 & $>3$\\
FIRST & 1400 &  0.083  &  1 & 811117 & 3 & 5 \\
NVSS  & 1400 &  0.75     &  2 & 1773484 & 4 & --- \\
SUMSS &  843  &  0.75     &  8 & 134870 & 5  &3 \\
MRC & 408 &  3.0      &  700 & 12141 & 6   & 7  \\
TXS  &  365  &  0.1      &  250 & 66841 & 7 & 5 \\
WISH & 325  &  0.9      &  10 & 90357 & 8 & 10\\
WENSS & 325  &  0.9      &  18 & 229420 & 9 & 6\\
MIYUN & 232  &  3.8      &  100 & 34426 & 10 & 5\\
4C  &  178  &  11.5     &  2000 & 4844 & 11 & 15 \\
MWA32 & 150 & 15 & .5 to 10 Jy & 1553 & 12 & 20\\
PAPER32 & 150 & 15 & 10 Jy & 486 & 13 & 20
\enddata
\tablecomments{{\bf References.} 
(1) \cite{Wright:1990p8680,Griffith:1994p8779};
(2) \cite{Otrupcek:1991p8780};
(3) \cite{White:1997p8645};
(4) \cite{Condon:1998p7986}; 
(5) \cite{Mauch:2003p8804};
(6) \cite{Large:1981p7798,Large:1991p7760};
(7) \cite{Douglas:1996p8840};
(8) \cite{DeBreuck:2002p8677};
(9) \cite{Rengelink:1997p8879};
(10) \cite{Zhang:1997p8880};
(11) \cite{Pilkington:1965p8882,Gower:1967p8886};
(12) \cite{Williams:2012p8768};
(13) \cite{Jacobs:2011p8438}
}
\end{deluxetable}

The construction of a catalog necessarily involves the compression and
omission of information, but in the context of the above goals, we can
ask three basic questions when comparing catalogs:
\begin{enumerate}
\item How were the flux scales established for each catalog, and are
  they consistent with each other?  This is a question about the
  average properties of the catalog fluxes, and does not imply that
  any particular source has an accurate flux.
\item Are the random errors in the source fluxes, relative to the
  fundamental flux scale, correctly described by the error bars
  presented?
\item Are there systematic effects, known or suspected, which are not
  reasonably described by the error bars given?
\end{enumerate}

Answering the first question requires establishing a certain source or
sources to use as references, and a method for comparing to them.
Ideally, a detailed model exists for the calibration sources,
including their spatial and spectral structure at the frequencies of
interest, as well as a model for their variability, if any.  A key
reference catalog for southern hemisphere low frequency radio sources
is the fan-beam survey with the Culgoora Circular Array\footnote{Normally referred to as the Culgoora Radio Heliograph (CRH), at nightfall the telescope became the Culgoora Circular Array (CCA)}
\citep[CCA;][]{Slee:1995p7541,Slee:1975p9320}.  The CCA produced the
so-called ``Culgoora'' catalog of fluxes at 80 and 160 MHz.  At 160
MHz, the CCA had 1.6\arcmin\ resolution and a narrow (1 MHz) bandwidth
\citep{Sheridan:1973p8795}.  The CCA catalog's flux scale is derived
from the CKL scale \citep{Conway:1963p9590}, as revised in \citet{Slee:1995p7541},
which is ultimately tied to the flux of Cassiopeia A.  The Culgoora
catalog was compiled from observations over the years 1970 - 1984.
Its status as the only low-frequency radio catalog in the southern
hemisphere has placed it a the center of the calibration schemes for
both PAPER and MWA, but it is well to keep in mind that is was very
different instrument than current EoR telescopes in terms of bandwidth
and resolution, and the Culgoora catalog lacks information on the
extent and spectral index of sources.

As for the second and third questions, we expect that the various
kinds of errors which can occur in reported fluxes to behave
differently according to their origin.  Errors resulting from random
noise are the simplest, and are at a value fixed by the local noise
level.  In a fractional sense, these errors are worst for the lowest
signal-to-noise sources, and indeed, for $S/N < 5$, reported fluxes
from blind catalogs tend to be systematically biased high due to
so-called Eddington bias \citep{Eddington:3p9023} (unless precautions
are taken.)  Most surveys at low frequencies are {\it not} dominated
by their random errors.  For example, the ongoing GMRT 150 MHz
survey\footnote{\url{http://tgss.ncra.tifr.res.in/}} reaches an RMS
noise $\sim8$~mJy beam$^{-1}$, but the flux scale accuracy is limited
by systematic errors to about 25\%.  
Errors due to source fitting, photometry, or {\sc clean}ing of a given
source can all be expected to scale in proportion to the source flux,
since these methods tend to over- or under-estimate by some fraction
of the flux, which means these produce a fixed fractional error.
Sources which are affected by the improperly convolved sidelobes of
another source can expect to have discrepancies in their recovered
flux which are uncorrelated with their flux level.  In addition to
errors introduced by the data reduction, other kinds of systematic
discrepancies between measurements may be introduced either by the
instrument or by natural processes.  On the instrument side, these
effects include an incorrect primary beam model, the presence of radio
frequency interference, or improper bandpass and source spectrum
calibration.  Physical processes include ionospheric variability,
interstellar medium scintillation, and intrinsic source variability.
Though most catalogs have only a limited model of these kinds of
errors folded into the listed error bars, systematic effects are often
discernible.


%


In this paper, we compare the recently published PAPER
\citep{Jacobs:2011p8438} and MWA \citep{Williams:2012p8768} catalogs.
Because the observations were made by these instruments within months
of each other, in overlapping portions of the sky, using very similar
configurations and bandwidths, we expect that disagreements between
sources due to time variation, spectral slope and confusion are
minimized.  Nevertheless, specific instrument differences including
modeling of the primary beam and \clean ing remain between the two
catalogs, as well as differing fields-of-view, noise depth, and
catalog construction method.  Our goal is to further understand the
origins of errors in published source fluxes of catalogs from EoR
surveys.  By limiting the data scope to only published data, we will
characterize the degree to which published catalogs provide all the
information necessary to reconstruct the sky model.  This will further
the overarching goal of refining our ability to reliably describe and
exchange sky models for the purposes of calibration and consistency
checks.

The outline of the paper is as follows.  The PAPER and MWA
observations are described in Section \ref{sec:data}.  Section
\ref{sec:comparison} compares the PSA32 and MWA32 catalogs in their
region of overlap and introduces a robust statistical comparison
method that uses a Markov Chain Monte Carlo (MCMC) algorithm to
compute the relative flux scale and its error.  Section
\ref{sec:internal_catalog_uncertainty} looks for systematic effects in
both data sets by internal comparison of the MWA data and comparison
of the PAPER data against the Culgoora catalog.
Section \ref{sec:Recommendations} summarizes the various errors identified, 
and concludes with recommendations for future EoR foreground
cataloging and results comparison efforts.

\section{The MWA32 and PSA32 Data Sets}
\label{sec:data}

The MWA in Western Australia and PAPER in South Africa are both
actively observing as their commissioning progresses. As part of the
EoR effort the observers are generating ``global sky models'', a key
component of which is a point source catalog.  First-look catalogs
using data taken during 2010 are now available from both instruments.
Relevant data about the two catalogs are listed in Table
\ref{tab:catalog_properties}.  Both instruments operated 32-antenna
arrays centered at an observing frequency near 150 MHz with similar
antenna layouts and bandwidth that result in an apparent resolution of
$\sim$15\arcmin.\footnote{Sources having more power at lower
frequencies can have an effective synthesized beam 35\arcmin\ wide
compared with brighter at the higher end of the bend which would have
an effective width of only 15\arcmin.}

\begin{deluxetable}{rrr} 
\tablecaption{Observation properties of the catalogs under study} 
\tablehead{
Telescope & PAPER & MWA}
\startdata
\label{tab:catalog_properties}
Resolution [arcmin]& 15  &  15\\
Bandwidth [MHz] & 60 & 92.16 \\
Center Frequency [MHz]& 145 & 154\\
Integration Time [minutes] & 30 & 40-200\tablenotemark{a} \\
Image plane RMS [Jy] & 2 & 0.2\\
 Lower flux limit [Jy] & 10 & 0.5 \\
Catalog method & Targets  & Blind  \\
Area covered [sq deg]&36000 &2600\\
Observation dates & May \& Sept 2010 & March 2010
\enddata
\tablenotetext{a}{Integration time varies between two the two facets, each of which is a drift scan which effectively spreads the integration time across the image.}
\end{deluxetable}

The PAPER data set \citep[][hereafter PSA32]{Jacobs:2011p8438}
consisted of observations on two nights separated by 3 months, using
32 single-polarization dipoles with 60 MHz of bandwidth centered at
145 MHz.  Both nights were used to make a mosaic covering the entire
sky with $\delta<10\arcdeg$.  The brightest two sources were used for
phase calibration and then filtered in delay-delay rate space
\citep{Parsons:2009p7859}.  The visibilities were imaged in ten minute
transit ``snapshots'' and then mosaiced into a single HEALPix
\citep{Gorski:2005p7667} image.  An image-based \clean
\citep{Hogbom:1974p7606} was performed on the brightest sources, but
most of the image was left un-\clean ed. For this reason the depth of
the catalog was kept to the brightest few sources in the sky. The
fluxes given in the PSA32 catalog are the peak flux within
30\arcmin\ of locations of catalog sources chosen from the Molonglo
Reference Catalog \citep[MRC;][]{Large:1981p7798,Large:1991p7760}. In
a selection designed to be complete at the minimum flux, it includes
all sources above 10 Jy as extrapolated to 150 MHz using the catalog
spectral index. The PSA32 fluxes compared to MRC and Culgoora showed a
similar range of variance about unity flux scale as the MRC and
Culgoora showed between themselves.

The MWA32 images were made from several nights of data in March 2010
from scans of two fields centered on RA 9h18m6s, Dec -12d05m45s (Hydra
A) and RA 10h20m0s, Dec -10d0m0s (EOR2).  
Imaging was performed in three 30 MHz bands which were averaged into
one 90 MHz wideband image on each field.  In this average the three
maps were weighted by a positive spectral index of 0.8 to compensate
for the average spectral index of -0.8. For sources with a spectral
index of -0.8, this will increase the perceived flux by 2.5\% as well
as slightly shrink the effective PSF by emphasizing higher frequency
data.  The images include both more integration time and more
snapshots, than the PSA32 observations, and were cleaned to a much
deeper level.  The MWA catalog sources were found blindly in this
wide-band image, without any catalog prior. Peaks having ${\rm SNR}>
3$, where the noise level is the average nearby image RMS, were fit
with two dimensional Gaussians.  The ${\rm SNR}=3$ sources range in
RMS from 167 mJy to 3 Jy, and 0.5 to 10 Jy in amplitude as the noise
varies across the map. The catalog lists the Gaussian amplitude of all
fits that converged, but not the sizes and orientations of the
Gaussians.  The derived fluxes were found to be within 30\% agreement
of the MRC predicted flux, which was then given as the data point
uncertainty.

\begin{figure}[htbp]
\includegraphics[width=0.9\columnwidth]{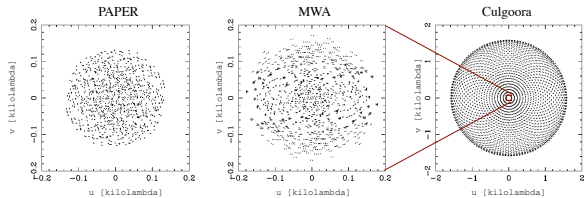}
\caption{\label{fig:uv} The uv sampling of PSA32 and MWA32 are very
  similar in scale and coverage density, with baselines between a few
  and 1000 meters, but are very different from larger instruments like
  the CRH/CCA (which made the majority of the southern hemisphere flux
  measurements at 150 MHz), whose shortest baseline was 100 m. For
  this reason we focus here on a comparison between PAPER and MWA.
  The uv coverage is shown at a single 150 MHz channel which is
  representative for the Culgoora 1 MHz passband. Both PAPER and MWA
  images were made over $\sim$100 MHz of bandwidth, and thus have
  $\sim$100 times more uniform uv coverage in multi-frequency
  synthesis images.}
\end{figure}

The PAPER flux scale was derived by calibrating each epoch to a single
Culgoora source, using 1422-297 for the May and 0521-365 for the
September data.  The calibration was effectively applied to the entire
image by the use of a primary beam model.  The MWA flux scale was
derived from an ensemble of sources with fluxes at 80 and 160 MHz from
the CCA, and 408 MHz from MRC, so the fluxes used by
\citet{Williams:2012p8768} were not precisely those of the CCA 160 MHz
catalog, though they are of course closely tied to them.
The use of Culgoora by both instruments to set a flux scale does not
of course allow us to address the absolute accuracy of the
measurements, which ultimately depends on the CKL flux scale.  The
applicability of the Culgoora fluxes is more generally subject to some
concern.  The narrow bandwidth of Culgoora and the lack of precise
spectral index information means that, integrated over $\sim$100 MHz
of bandwidth, a source with a spectral index $\alpha\sim-1$ will
appear 5\% brighter than a narrow spectrum measurement.  Large scale
structure invisible to the CCA could substantially boost the flux for
resolved sources observed dense aperture arrays like the MWA or
PAPER. As shown in Figure \ref{fig:uv} the MWA and PAPER 32 antenna
arrays are much more compact and have little overlap with the long
baselines of the CCA.  The images shows the narrow-band $uv$ coverage;
in fact, PAPER and MWA cover nearly 100 times as much $uv$ space in a
multi-frequency synthesis image.

\begin{figure*}[htb]
\begin{center}
\includegraphics[width=0.95\textwidth]{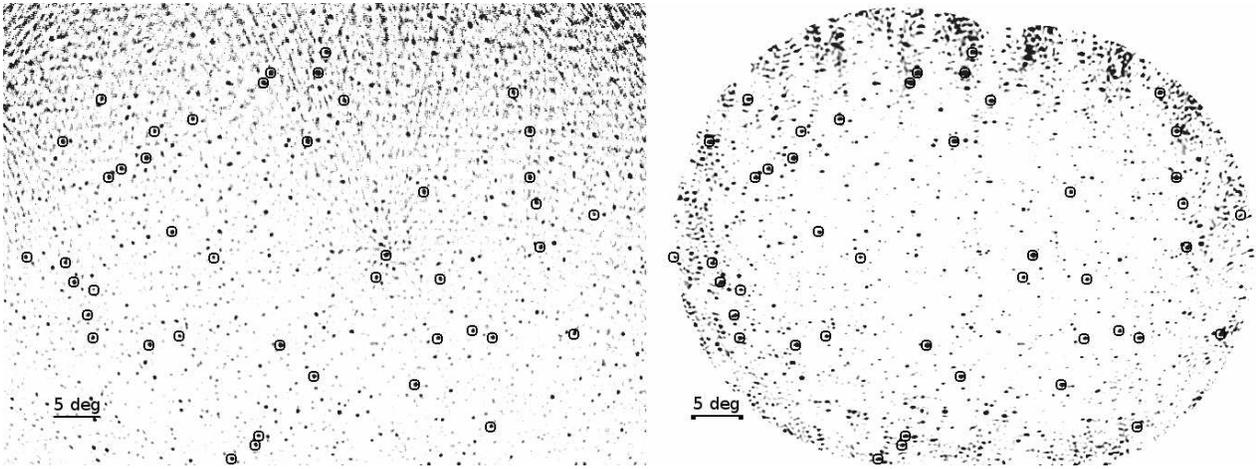}
\caption{A side-by-side comparison of PSA32 (left) and MWA32 (right)
  images under study here. The PAPER image is a mosaic of several
  snapshots that have been weakly cleaned. The bright side-lobes are
  due to residual Hydra A flux remaining after delay-delay rate
  filtering.  The MWA mosaic is formed by averaging the two facets in
  Williams et al (2012) with a 10\arcdeg-wide gaussian weight. The MWA
  images are composed of several drift scans and, while having a
  variable noise across the image do not have a simple corresponding
  set of primary beam weights. Sources found in both catalogs are
  black circles. Both images are centered on RA 9h45m-10d, 70 degrees
  wide by 50 degrees tall, and a pixel size of 3\arcmin. The color
  scale is set so that 90\% of the flux scale is black. }
\label{fig:mosaics}
\end{center}
\end{figure*}

Despite the high level of similarity between the two data sets, there
are still important differences which should be carefully noted.
Probably most importantly, are the differences in image depth and
area.  The PSA32 images incorporate data from many different pointings
to smoothly map the sky; signal-to-noise is relatively constant across
the image, but due to limited deconvolution the dynamic range is
lower. The MWA images are more deeply deconvolved but limited in
extent. The difference in SNR between the middle and edges is
pronounced and comparable to the areas of the PSA32 map dominated by
side-lobes.  Figure \ref{fig:mosaics} directly compares the images of
the overlap region from both instruments.


In addition, spectral slope across the wide $\sim$80 MHz bandwidths
used by PAPER and MWA could also be a source of intrinsic measurement
difference. The images used to build both catalogs incorporated data
across the band in a multi-frequency synthesis and thus are unable to
directly measure spectral index. The bandwidths are different by 30\%
which, for sources with large spectral slope\footnote{Most radio
  sources in this band have power law spectra $S(\nu) =
  \left(\frac{\nu}{\nu_0}\right)^\alpha$. The average spectral index
  for radio sources in this band is $-0.8$.} will result in slightly
different spectral averages.  Furthermore, the MWA32 sub bands were
weighted by the typical spectral index of $\alpha = -0.8$, while the
PAPER spectrum was not.  This will cause most sources to be on average
5\% brighter for MWA than for PAPER, additional spectral variations
between sources will introduce another $\sim$1\% variation around this
number.

 \section{Flux Scale Comparison Between Instruments}
\label{sec:comparison}

Each catalog provides a list of sources, each with a flux and flux
uncertainty. The PSA32 catalog lists peak flux and surrounding rms,
while the MWA32 catalog lists fitted flux and fractional error,
assumed to be constant at 30\%.  For the purposes of the following
analysis, we assume these errors correctly describe the instrumental
uncertainties.  This question is explored further in Section
\ref{sec:internal_catalog_uncertainty}.

In the region of overlap between the two surveys,
there are 60 MWA entries within 30\arcmin\ of 41 PSA32 sources. Of
these 41 PSA32 sources, 13 have multiple MWA components while the rest
are 1:1 matches.  In the case of multiple component matches, we pair
sources with the highest flux. Images of the regions under comparison
along with markers for the 41 overlapping sources are shown in Figure
\ref{fig:mosaics}.

\begin{figure*}[htb]
\begin{center}
\includegraphics[width=0.45\textwidth]{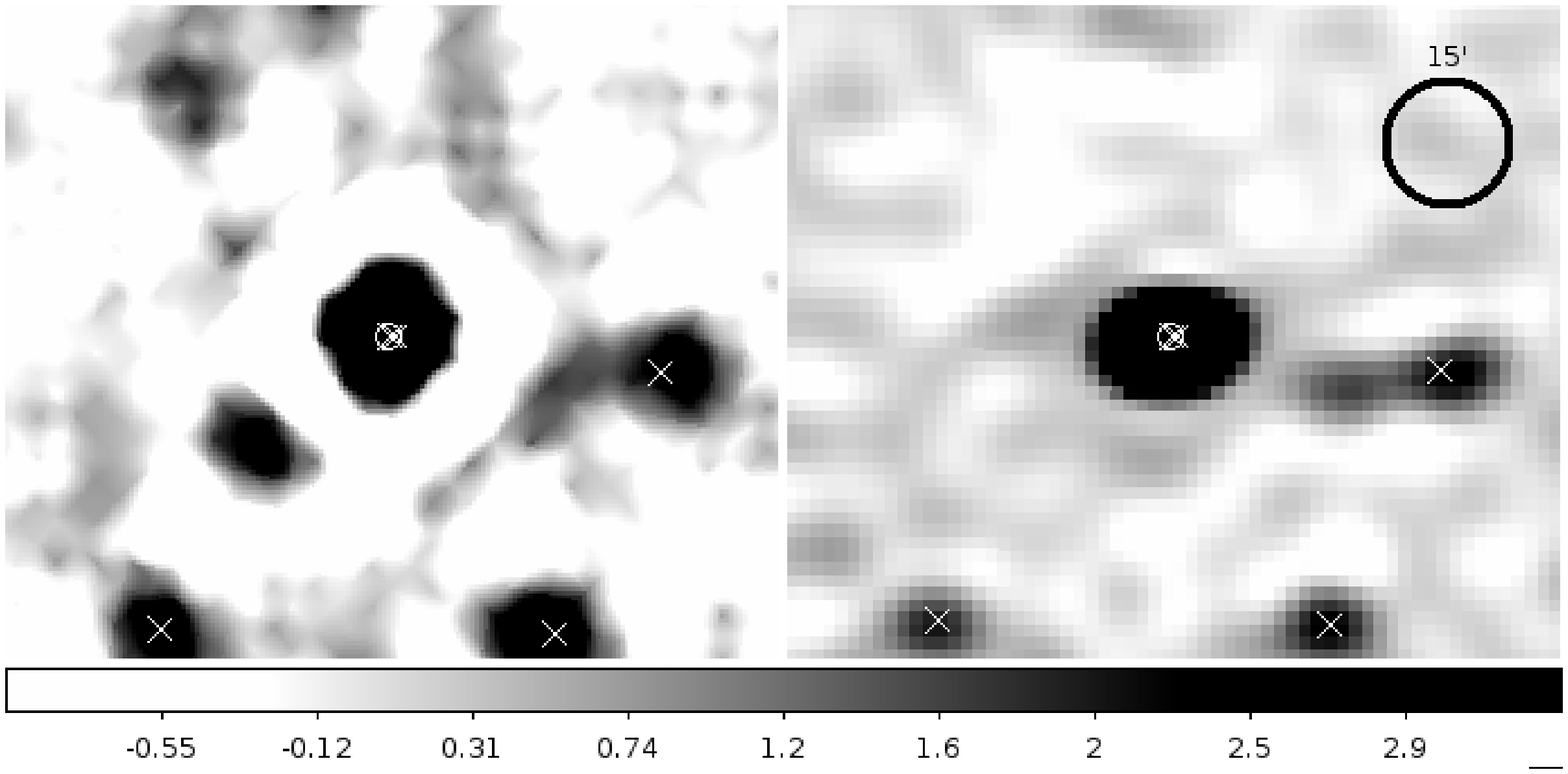}
\includegraphics[width=0.45\textwidth]{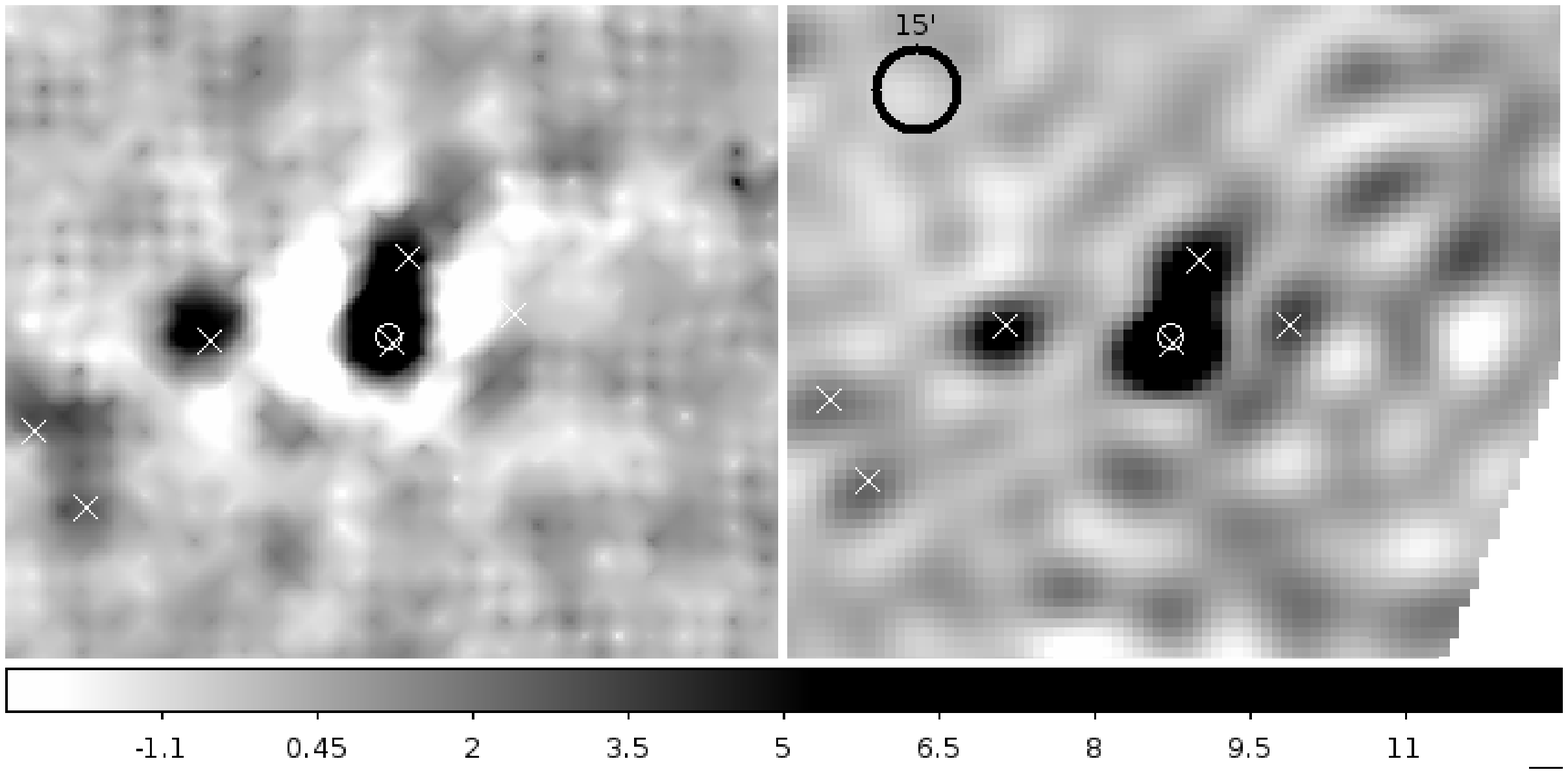}
\caption{A side-by-side comparison of two previously known sources
  (L:0859-257, R:0745-191) extracted from the mosaics in
Figure  \ref{fig:mosaics}. For each source, PAPER is on the left and MWA on
  the right with MWA32 catalog sources marked with an X; PSA32 listed the position and amplitude of the peak within 30\arcmin of the image center.
The left source
  provides examples of errors from both instruments.  The MWA catalog
  lists two sources separated by 1.4\arcmin, or 10\% of a synthesized beam,
  which were even given the same truncated J2000 name. Meanwhile, the
  PAPER image, having not been \clean ed to this flux level, has larger
  side-lobes. Together these effects contribute to a 180\% flux scale
  between the two (28 Jy for PAPER, (43+6) Jy for MWA). However,
  differences in deconvolution do not preclude an accurate comparison
  as shown on the right, where a source has multiple confused
  components yet the PAPER flux is within 17\% of the MWA flux.}
\label{fig:image_detail}
\end{center}
\end{figure*}

Two of these sources provide instructive examples. Figure
\ref{fig:image_detail} shows the PAPER and MWA images for two of the
brightest sources which are listed in both catalogs and have multiple MWA
components within 30\arcmin of a single PAPER source. The first,
J0859-257, demonstrates the importance of both \clean and cataloging
method. The MWA32 catalog lists two sources in virtually the same
location. (They are separated by 1.4' or 1/10 of a synthesized beam
and were given the same truncated J2000 name.). Meanwhile, the PAPER
image which was not cleaned to this level has deep side-lobes and
excess flux not visible in the MWA. Together these effects contribute
to a 180\% flux difference between the two (28 Jy for PAPER, (43+6) Jy
for MWA). The second source shown, J0745-191, is a classic example of
resolution confusion, two sources whose point spread functions
significantly overlap. Despite this, the two instruments agree on the
brighter flux to 17\%.

%

%

Having obtained a list of corresponding sources, we wish to ask
whether the two instruments produce measurements that are consistent
with being on the same flux scale, given their reported errors.
We thus compute the likelihood that the PSA32 fluxes $S_P$ are related
to the MWA fluxes $S_M$ by a simple linear fit, with deviations from
this relation due solely to random errors as described by both
instruments' error bars.  The likelihood function is given by
\citet{Hogg:2010p8759}
in their \S 7. 
We implement a Markov Chain Monte Carlo sampler to sample the
posterior probability.  
At each step of the Markov chain we compute the error and distance of
each point as projected orthogonally to the current direction of the
line. These differences and errors are then used to form a Gaussian
likelihood. The free parameters are the slope and the offset of the
flux-flux line.

\begin{figure}[htb]
\begin{center}
\includegraphics[width=\columnwidth]{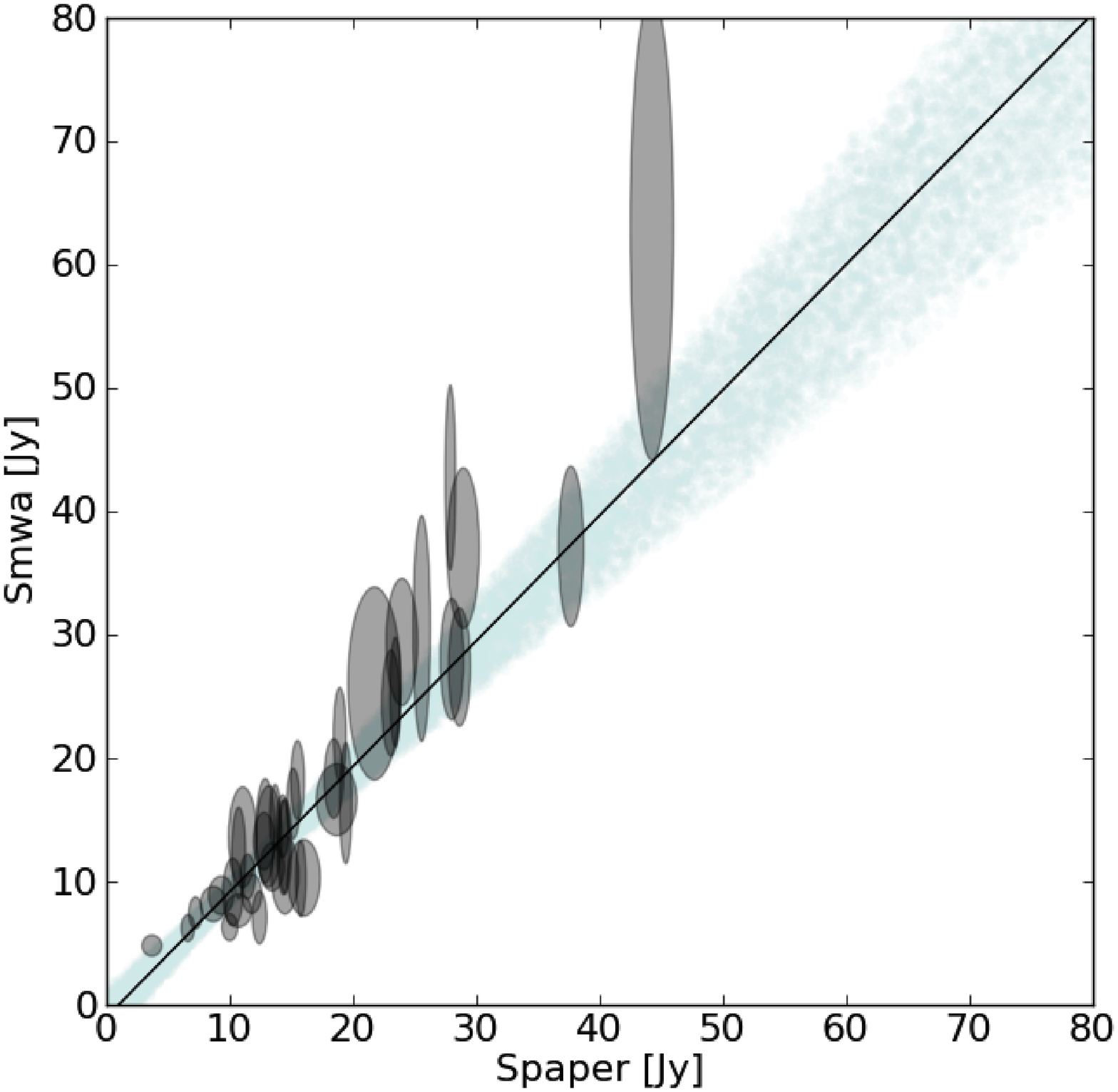}
\caption{Fitting a linear relationship between MWA and PSA32 in the
  presence of error bars.  The PSA32 errors are image plane RMS in an
  annulus around the source, while the MWA error bars are fractional
  between 30 and 80\%, depending on distance from image center. The
  line represents the peak of the posterior and the blue region
  indicates $1\sigma$ confidence.  }
\label{fig:ff}
\end{center}
\end{figure}

\begin{figure}[htb]
\begin{center}
\includegraphics[width=0.9\columnwidth]{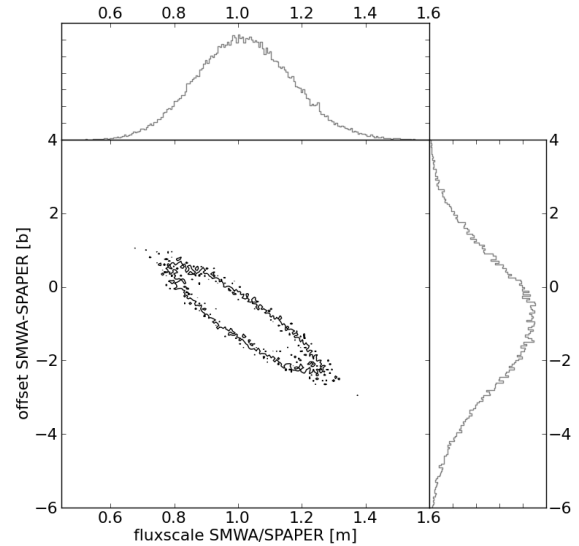}
\caption{The posterior probability distribution of the PAPER/MWA flux calibration. The output of the flux relationship fit is a series of
  samples of the model parameters, slope (m) and intercept (b).  The
  occupation number of each m,b value is a sample of the posterior
  probability and the projection down to either variable gives the
  marginalized distribution. The histograms on the sides give the
  marginalized likelihood. The marginalized flux-scale or slope
  is analogous to the distribution of flux scales used in
  \cite{Jacobs:2011p8438}.  The peak probability occurs at flux scale
  of unity and intercept of -0.5, the contour shown encloses the
  solutions having 76\% probability. The slope of the probability
  distribution is steep; 95\% probability density contours were not
  significantly different enough to be over plotted. The marginalized
  slope posterior, labeled as "flux scale" to which it is roughly
  analogous, reaches the 76\% level at 0.8 and 1.2 indicating that the
  flux scale is correct to within 20\%.}
\label{fig:m_b_compare}
\end{center}
\end{figure}

The posterior probability distribution 
is shown in Figure \ref{fig:m_b_compare}. 
%
The most likely flux relationship occurs at the peak of the posterior
(shown in Figure \ref{fig:ff}), and the confidence interval is defined
as the contours of the posterior sampling. Marginalizing over the flux
offset, we find a distribution of flux scales which peaks at 1.05 and
has a 73\% confidence limit of 0.8 to 1.19, or 20\% at $1\sigma$. The
peak position is consistent with the offset due to the small spectral
index correction in the construction of the MWA32 wideband images
and is consistent with MWA32 and PSA32 sharing the same flux scale.
It should be emphasized that this is a more robust and correct
determination of relative agreement between catalogs than either the
flux-ratio histogram method implemented by \citet{Jacobs:2011p8438} or the
average flux ratio of \citet{Williams:2012p8768}.


%

\section{Systematic Effects in the Catalogs}
\label{sec:internal_catalog_uncertainty}

Both the PSA32 and MWA32 source catalog errors are almost certainly
{\it not} dominated by thermal noise.  To assess the origin of errors,
it is necessary now to turn to possible sources of systematic errors,
and, for this purpose, it is desirable to have a reference to compare
against.  Since the MWA32 catalog is derived from sources found in two
facets, it is possible to use intercomparison between the two facets
as a diagnostic of systematic errors.
While in principle a similar approach could be used for the PAPER
images, the individual PAPER snapshots were of a limited
signal-to-noise, and thus intercomparison is not very meaningful.
For this reason the individual facets were neither published nor
included in this study.  Thus for PAPER, we look for systematic errors
by comparing against the CCA catalog.


To simplify the analysis we will compare the peak fluxes, rather than
the Gaussian fits used in \citet{Williams:2012p8768}.  This also
simplifies the comparison to the PSA32 catalog (Section
\ref{sec:comparison}), which also used peak fluxes. To test the actual
amount of flux error when using the two methods we compare the MWA32
peak fluxes with the fit fluxes listed in the catalog.  The amount of
disagreement ranges from a median of $<1\%$ in the Hydra A field to
8\% in the EOR2 field. As we will see, this error is much smaller than
other effects we will identify.

Occasionally, several MWA
sources were closer together than 30\arcmin\, causing the peak finder
to sometimes find duplicate flux measurements. After eliminating
$\sim$10 sources within 30\arcmin\, of each other, we compute the
median and rms facet to facet fractional difference.

\begin{figure}[htbp]
\begin{center}
\includegraphics[width=0.9\columnwidth]{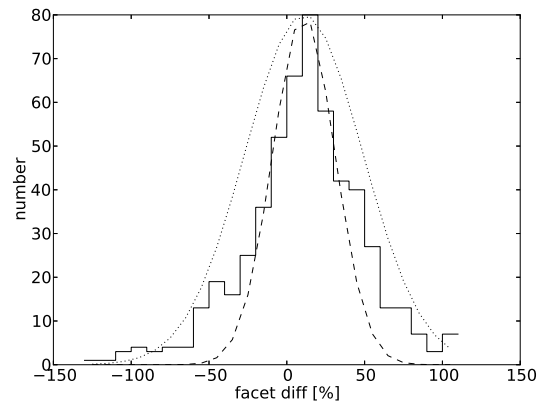}
\caption{\label{fig:err_hist} The two overlapping MWA facets provide
  an opportunity to examine the sources of errors. Here we examine the
  distribution of fractional difference in the facet flux of MWA32
  catalog sources. The distribution is non-gaussian which causes the
  gaussian fit (dotted) to clearly over-estimate the amount of error
  at 37\% compared with the 20\% error model found by comparing with
  PAPER (dashed). In these images most of the sources with 50\%$+$
  error appear to be the result of an imperfect primary beam model
  correction (c.f. Fig. \ref{fig:flux_err_dist_ra} showing this error
  vs RA and Fig \ref{fig:beam_differences} giving an explanation for
  the shape of that relation).  }
\end{center}
\end{figure}

In this large sample of 539 sources, the distribution of the
fractional errors is peaked around 16\% but extends beyond 100\%, a
state reflected in its RMS of 39.9\% and median value of 13\%.  The
distribution of the errors is shown in Figure \ref{fig:err_hist}. The
best fit histogram has a width of 37\% though a width of 20\% seems to
better reflect the center of the distribution, which, as we will see
below suggests that the errors are non-gaussian and most likely systematic.
  
Though many sources are visible in both facets only the subset found
in the primary field of view\footnote{The actual effective beam will
  be complicated by the inclusion of several pointings and bands, all
  of which have measurably different patterns.  This sample, which
  includes only the published maps and the known primary beam size
  probably best describes the uncertainty in the MWA32 catalog.}
(26\arcdeg FWHM @ 190MHz) have comparable instrumental error. Indeed,
the median uncertainty of these 63 sources very similar to the larger
sample at 16\%, but the RMS is much smaller at only 29\%.

%
%

\subsection{Errors Due to Primary Beam}

\begin{figure}[htbp]
\begin{center}
\includegraphics[width=0.9\columnwidth]{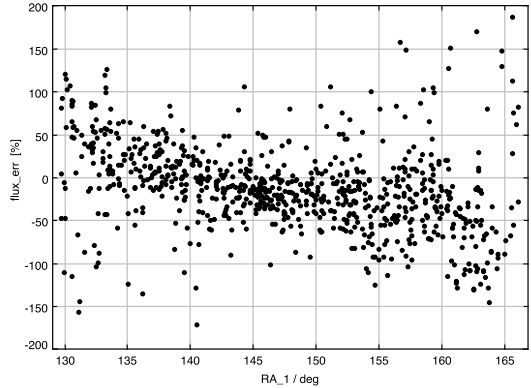}
\caption{\label{fig:flux_err_dist_ra} Fractional difference between
  peak fluxes in the two overlapping MWA32 facets as a function of
  Right Ascension. No trend is observed in the Declination direction.
  The fractional difference near the middle, where the facet overlap
  is best, averages around 20\% and rises to over 100\% at the
  periphery. The shape is similar to what one would expect from use of
  an inaccurate primary beam model, a problem endemic to both PAPER and
  MWA. See Figure \ref{fig:beam_differences} for a cartoon explanation.}
\end{center}
\end{figure}

Meanwhile, the opposite is true of flux difference versus
Right Ascension, as is shown in Figure \ref{fig:flux_err_dist_ra}.
Sources above 1 Jy show a clear linear trend in flux difference with
Right Ascension, changing by as much as 200\% over 25 degrees of
longitude. No trend is observed in the Declination direction.

\begin{figure}[htbp]
\begin{center}
\includegraphics[width=0.9\columnwidth]{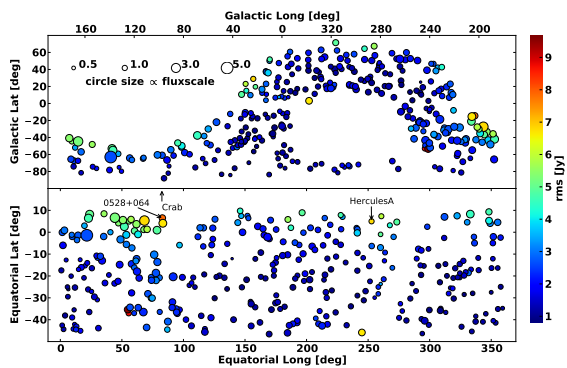}
\caption{\label{fig:PAPER_fluxscale} Spatial distribution of
  PSA32/Molonglo Reference Catalog flux ratio and local PSA32 rms in
  Galactic (top) and Equatorial (bottom) coordinates (from
  \cite{Jacobs:2011p8894}). Point size indicates flux scale as shown
  in inset key, local image rms is related by color. The area of high
  flux-scale appears to be correlated with high rms in upper
  latitudes, particularly near bright sources far from pointing
  center.  Though the error is not strictly linear with distance from
  the suspected source of side-lobes, inspection of the image suggests
  that insufficient deconvolution of Hercules A and the Crab is to
  blame. }

\end{center}
\end{figure}

\begin{figure}[htbp]
\begin{center}
\includegraphics[width=0.9\columnwidth]{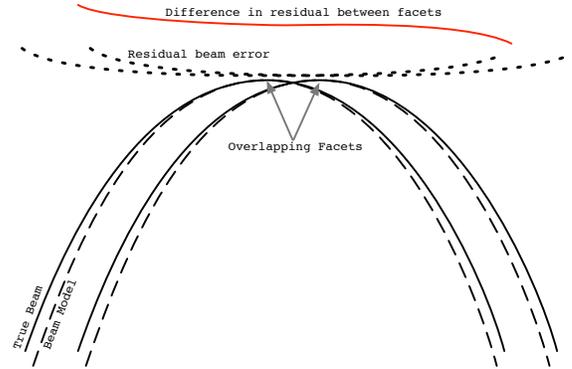}
\caption{\label{fig:beam_differences} A cartoon of a 90\arcdeg azimuth
  (East-West) cut through two adjacent primary beams of any wide field
  telescope to compare with the systemic difference shown in Figure \ref{fig:flux_err_dist_ra}.  When the model of the primary beam (solid black) is
  applied in place of the true model (dashed) the error is manifested
  as a characteristic flux-scale that varies with position (dashed
  lines). When two pointings are differenced, the errors on the
  opposing edges will have opposite signs. The affect will be most
  pronounced when comparing sources occurring at the extremes of both
  beams along the axis bisecting both facets (here Right Ascension).}

\end{center}
\end{figure}

A second systematic affect was apparent in the faint, un\clean ed
sources $<5$~Jy of the MWA32 facets: a distinct systematic, monotonic trend in facet
disagreement (Figure \ref{fig:flux_err_dist_ra}). The RA dependence of
the disagreement is consistent with the expected difference between
two facets with similar scale beam errors as illustrated in Figure
\ref{fig:beam_differences}. The true flux of each facet image is
estimated by dividing the perceived flux by a model of the primary
beam. The models used for both MWA and PAPER are based on
simulations. When the model does not match reality the flux scale will
incorrectly be seen to increase or decrease uniformly towards the beam
edges.  When two pointings are differenced, the errors on the opposing
edges will have opposite signs. The scale of the error, $\sim 40$\% at
field of view edges, is consistent with the tests of MWA antenna tiles
in an anechoic chamber at Lincoln Labs, where the MWA tile responses
were found to differ from the model by 1.34 dB (36\%) at
15\arcdeg\ from zenith \citep{Williams:2012p8895}.

MWA primary beams will be holographically mapped and calibrated during
commissioning of the final array configuration. This data was not yet
available for the MWA32 catalog.  Because the MWA images from this
study used several tile aperture pointings and the more limited
theoretical holographic model, the uncertainty in the beam model was
large. Other experiments by \cite{Bernardi:2012p9020} observing in
drift scan mode using only a single, well characterized, pointing,
were able to find closer flux/catalog agreement. Methods that utilize
the holographic beam in the deconvolution process are now being tested
that will significantly reduce this systematic
\citep{Morales:2009p8566,Sullivan:2012p9457,Tasse:2012p9463}.

\subsection{Errors Due to \clean\ Algorithm}

As we saw when comparing images in Fig \ref{fig:image_detail}, different levels of deconvolution affect the degree to which the PAPER and MWA images agree.  This affect is also noticeable when comparing the two MWA facets which were cleaned independently.

\begin{figure}[htbp]
\begin{center}
\includegraphics[width=0.9\columnwidth]{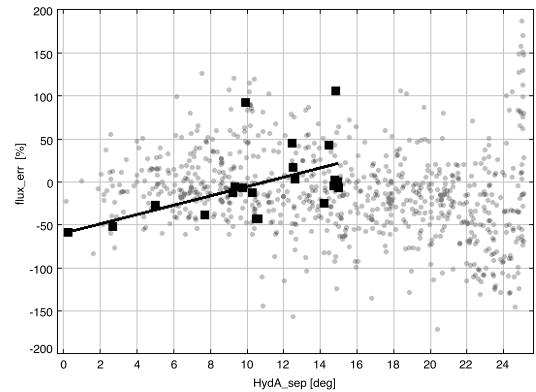}
\caption{\label{fig:flux_err_dist_hyda} Fractional difference between
  peak fluxes in two overlapping MWA32 facets versus distance from
  Hydra A, which is 7 times brighter than the next brightest source in
  either image.
The error in bright
  sources ($>1$Jy, black squares) generally tracks that of 
the full set of sources (gray dots), but the bright sources nearest Hydra A show a depression of their flux consistent with sitting in a negative sidelobe of Hydra A. The black line is not a formal fit but shows the systematic suppression of bright sources $< 8$ degrees from Hydra A.
}
\end{center}
\end{figure}

In this limited selection, systematic differences errors are less obvious. One that is most suggestive is a possible linear increase in error with proximity to Hydra A shown in Figure 10. Hydra A is 7 times brighter than the next brightest source. 
During the first \clean\ iterations
the model will only contain Hydra A.  When \clean\ begins to model flux
at the level corresponding to the next brightest sources it must
decide how to divide up fluxes of nearby sources whose side-lobes
significantly overlap. If it divides incorrectly (putting the flux of
one source into another) \clean\ enters a false minimum from which it
cannot escape.  The result will be that models of sources near very
bright sources will be more corrupted.  The images were \clean ed to 1\%
of the peak flux, or about 4 Jy for EoR2 and 6 Jy for the Hydra A
field . The fact that the error does not affect sources below 5 Jy
suggests that the error is related to a \clean\ converging on a false
minimum.

Deconvolution and primary beam errors illustrated by the MWA32 data
are present in varying degrees in the PSA32 data as well. The PAPER
images are not \clean ed as deeply as the MWA32 images. However, as we see in Figure \ref{fig:PAPER_fluxscale}
which shows the ratio of PAPER to catalog values, when compared to
other catalogs, the largest errors were found to cluster near bright
sources beyond the imaged region at low elevation in the primary beam
\citep{Jacobs:2011p8894}. The errors did not increase with distance and
appear to be due to side-lobes from the sources indicated in the
figure. Recent analysis of measured source tracks has found the PAPER
beam to be accurate to between 10\% and 15\%, though sources can have
individual errors of 20\% or occasionally more \citep{Pober:2012p8800}.

\section{Conclusions and Recommendations}
\label{sec:Recommendations}

We summarize our conclusions as follows:
\begin{enumerate}

\item The PSA-32 and MWA-32T catalogs are on the same flux scale,
  consistent with their stated errors (flux agreement of 20\% at a
  probability of 0.76).

\item Both PSA-32 and MWA-32T catalogs show evidence for systematic
  errors in the fluxes of sources near bright sources, the likely
  explanation for which is errors in \clean ing the bright sources.

\item The MWA-32T catalog shows evidence for a systematic flux error
  of sources as a function of RA likely due to an error in the primary
  beam model combined with the mosaicking of facets along the RA
  direction.  Due to its construction from a number of overlapping
  facets along RA, the PSA-32 catalog does not show a similar
  artifact.

\end{enumerate}

%

We summarize the sources of error in the three sets of flux
measurements (two MWA32 facets, 1 PAPER mosaic) into several
categories, outlined in Table \ref{tab:ErrorBudget}.  Types of errors
as deduced from the MWA facet analysis are given in the upper part of
the table, whereas intercomparisons between the two catalogs are given
below the dividing line.

\begin{table}[hbt]
\begin{center}
\caption{Error Budget}
\label{tab:ErrorBudget}
\begin{tabular}{lcc}
\hline
\hline
Source of Error & Fractional Error & Refer to \\
\hline
Flux measurement (peak vs fit) & 4.5\% & \S\ref{sec:comparison} \\
Primary beam  & 25\% & Figure \ref{fig:flux_err_dist_ra}\\
Edge of beam & 100\% & Figure \ref{fig:flux_err_dist_ra}\\
\clean\ of bright sources & 50\% & Figure \ref{fig:flux_err_dist_hyda} \\
\hline
Theoretical bandwidth mismatch & 5\% &\S\ref{sec:data}\\
Actual difference between telescopes & 20\% & Figure \ref{fig:m_b_compare}\\

\end{tabular}
\end{center}
\end{table}

%

All EoR telescopes must demonstrate the ability to make reliable and
repeatable measurements. Employing the lessons learned in this early
stage we can
summarize the implications of Table \ref{tab:ErrorBudget} for
improvements necessary for EoR experiments are as follows:

\begin{enumerate}

\item \label{FluxScaleImplications} Flux scale is currently not accurate to better than 20\%. 
 This implies a $\sim40\%$ uncertainty in
  the $\Delta^2$ power spectrum.  This is most likely due to to
  primary beam uncertainties in transferring fluxes between
  calibrators, and also due to \clean\ uncertainties.

\item \label{SkyModelImplications} The precision of the sky model is
  sufficient to accurately subtract $\sim80\%$ of bright foreground sources,
  which is a significant distance from the 0.25\% requirement to be
  able to subtract sources and work within the EoR ``wedge''.  Future work
  should be able to improve on this dramatically, though it is not
  obvious that this two order of magnitude requirement can be reached.

\item \label{CLEANImplications} The \clean\ algorithm introduces correlated
errors between sources.  Catalogs should include information about the 
degree of correlation. This information would then inform the comparison
 likelihood model. 

\item \label{PrimaryBeamImplications} Work towards improving primary beam accuracy
 is of utmost
  importance for both experiments and for EoR measurements generally,
  as for polarization \citep{Moore:2013p9621}, image reconstruction and fully
  holographic imaging
  \citep{Sullivan:2012p9457} and is also currently the limiting factor in the
  accuracy of the catalogs.

\end{enumerate}

To address implication \ref{FluxScaleImplications}, we recommend establishment
of system of reference sources with detailed and repeated measurements by both
instruments.
We should note
that the only reason the flux relationship fit converges on a single stationary
gaussian-like probability distribution is the 30\% fractional error
bar listed in the MWA32. This large fractional error was designed to
match the approximate scale of deviation from Culgoora values and
appears consistent with the facet comparison analysis above (Figure
\ref{fig:err_hist}).  The significance of this comparison is in the
successful application of a new method for comparing catalogs. The
MCMC likelihood algorithm allows the addition of more detailed error
models that take into position and flux dependent errors like those
described above.  For the reasons noted in Section
\ref{sec:comparison}, inter-catalog comparisons should take into
account both the quoted errors, and quote the resulting range of model
parameters which could relate the two. It should be noted that the
probabilistic method used to relate PSA32 to MWA32 could be extended
to take into account a more detailed, non-gaussian, error model, and
in principle, can also be used to assess the correctness of the
individual object error bars from either catalog, with the addition of
a likelihood for the errors. Extra catalog meta data, such as 
the correlation between measurements, as suggested in 
number \ref{CLEANImplications}, could also be folded into the likelihood 
model. This is a subject for future work.

Regarding point \ref{CLEANImplications},
\clean\ incorporates little prior knowledge into its
result. This is a good choice for narrow field of view instruments
observing an unknown sky.  But wide field of view deconvolution always
encompasses many oft-measured sources.  Future deconvolution efforts should incorporate known
fluxes as prior data. One example of a method which could
incorporates priors in this way is the Fast Holographic Deconvolution
algorithm \citep{Sullivan:2012p9457} which provides a faster forward
model suitable for building a likelihood-based approach.


Of all the observed systematics, the beam model error is the
largest, making it clear that more effort must be devoted to measuring the
primary beam.  We note that in the case of the MWA, the beam error was
discernible because two deep, independently imaged facets happened to
overlap each other, allowing comparison of many dim sources.  This
suggests 1) that the images used to generate a catalog should be
published along with the list of source fluxes (both PAPER and MWA
images available only ``on request'') and 2) that surveys should be
arranged so that each source measurement is repeated at differing hour
angles, observing it at different points in the primary antenna beam.

\section{Acknowledgements}

This work makes use of the Topcat catalog
program\footnote{\cite{Taylor:2005p8683}
  \url{http://www.starlink.ac.uk/topcat/}} and the ``MCMC Hammer"
emcee python library\footnote{\cite{ForemanMackey:2012p8684}
  \url{http://danfm.ca/emcee/}} .

\end{document}